\documentclass[twocolumn]{aastex631}
\usepackage{CJK}
\usepackage{amsfonts}
\usepackage{amsmath}
\usepackage{mathrsfs}
\usepackage{epsfig}
\usepackage{subfigure}

\def\cd{\,\mathrm{c/d}}

\def\raa{Res.\ Astron.\ Astrophys.\ }

\shorttitle{Intricate Symphony in HADS}
\shortauthors{J.-S. Niu \& H.-F. Xue}

\begin{document}
\begin{CJK*}{UTF8}{gbsn}

\title{Unveiling the Intricate Symphony of Nonlinear Pulsation Mode Interactions in High-Amplitude $\delta$ Scuti Stars}

\correspondingauthor{Jia-Shu Niu}
\email{jsniu@sxu.edu.cn}

\author[0000-0001-5232-9500]{Jia-Shu Niu (牛家树)}
\affil{Institute of Theoretical Physics, Shanxi University, Taiyuan 030006, China}
\affil{State Key Laboratory of Quantum Optics and Quantum Optics Devices, Shanxi University, Taiyuan 030006, China}

\author[0000-0001-6027-4562]{Hui-Fang Xue (薛会芳)}
\affil{Department of Physics, Taiyuan Normal University, Jinzhong 030619, China}
\affil{Institute of Computational and Applied Physics, Taiyuan Normal University, Jinzhong 030619, China}


\begin{abstract}
People can diagnose the interiors of stars by sensing their pulsations. Pulsation modes, which are determined by the internal state and structure of a star, are typically considered stable over short timescales. These independent pulsation modes have been used in asteroseismology to reconstruct the interior structure of stars. Here, we report the discovery of peculiar pulsation mode interaction details in the high-amplitude $\delta$ Scuti star KIC 6382916 (J19480292+4146558), challenging the reliability of independent pulsation modes as indicators of the star's internal structure.
Through analysis of archival data, we found distinct variations in amplitudes and frequencies of three independent pulsation modes and their harmonics/combinations over approximately 20 days. These variations can reach amplitudes of about 140\% and frequency variations of about 12\%. Correlation analysis of amplitude and frequency variations revealed additional pulsation mode interaction details and patterns. Notably, our findings regarding the phenomena related to harmonics of independent pulsation modes challenge the traditional understanding in this area. These discoveries serve as cornerstones for future research and advance nonlinear asteroseismology.
\end{abstract}

\section{Introduction}

$\delta$ Scuti stars belong to the class of short-period pulsating variable stars, with periods ranging from 15 minutes to 8 hours and spectral classes A-F. They can be found at the intersection of the main sequence (MS) and the lower section of the classical Cepheid instability strip on the Hertzsprung-Russell diagram. The pulsations in $\delta$ Scuti stars are self-excited by the $\kappa$ mechanism, which arises from the partial ionization of helium in the outer layers \citep{Kallinger2008,Handler2009,Guenther2009,Uytterhoeven2011,Holdsworth2014,Steindl2022}.

High-amplitude $\delta$ Scuti stars (HADS) constitute a subclass of $\delta$ Scuti stars with relatively larger amplitudes ($\Delta V \geq 0_{\cdot}^{m}1$) and slower rotation speeds ($v \sin i \le 30\ \mathrm{km/s}$) in most cases. However, as more samples of HADS have been accumulated, these classical criteria have become less clear (see e.g., \citet{Balona2012}). Most HADS pulsate with one or two radial pulsation modes \citep{Niu2013,Niu2017,Bowman2021,Daszynska2022}, while some exhibit three radial pulsation modes \citep{Wils2008,Niu2022,Xue2023} or even non-radial pulsation modes \citep{Poretti2011,Xue2020}.

For a long time, the stability of the amplitudes and frequencies of HADS has been assumed based on the limited quality of observational data. However, recent advancements in continuous time-series photometric data from space telescopes, such as {\it Kepler} and {\it TESS}, have presented an opportunity to explore the variations of pulsation modes over several years. The analysis of high-quality data has revealed pulsation mode variations (including amplitudes, frequencies, and phases) in various types of stars, including RR Lyrae stars \citep{Carrell2021}, $\delta$ Scuti stars \citep{Murphy2012,Breger2014,Bowman2016}, pulsating hot B subdwarf stars \citep{Zong2016_sdb,Zong2018}, pulsating DB white dwarf stars \citep{Zong2016_wd}, and slowly pulsating B stars \citep{VanBeeck2021}. These variations not only provide additional information to understand pulsation modes themselves (such as the excitation and selection mechanisms, and mode interactions), but also shed light on stellar structures and evolution. Despite HADS being primarily pulsating in low-order radial p-modes (fundamental, first overtone, or second overtone modes), their pulsation modes also exhibit distinct variations over several years. Furthermore, the determination of linear period variation rates \citep{Bowman2021,Niu2022} and even quadratic amplitude variation rates \citep{Niu2022} from these photometric data offer further insights to improve current stellar evolution and pulsation theories.

KIC 6382916 (J19480292+4146558), also known as GSC 03144-595, is an HADS that was extensively monitored by the {\it Kepler} space telescope. The light curves of this star show three independent pulsation modes. Initially, these three modes were identified as non-radial pulsation modes with $l=1$ by \citet{Ulusoy2013}. Later, \citet{Mow2016} identified that these modes are actually the fundamental, first overtone, and second overtone radial p-modes. 
Recently, \citet{Niu2022} conducted a comprehensive study on the pulsation mode variations of KIC 6382916 over a period of about 4 years. Through a comparison between observations and theoretical calculations, they identified the first and second pulsation modes, labeled as $f_0$ and $f_1$, as the fundamental and first overtone modes, respectively. The third pulsation mode, labeled as $f_2$, was found to be the non-radial part of a resonating integration mode (RI mode), generated by the resonance between a radial p-mode and a nonradial p-g mixed mode.
Furthermore, \citet{Niu2022} discovered that almost all the $f_2$-related combinations have partners with frequencies approximately $\Delta \omega = 0.0815\ \cd$ away from them in the frequency domain. These pairs follow the frequency relation $f_2 - \Delta \omega = 2f_1 - f_0$.

In this study, we extracted the same pulsation modes as \citet{Niu2022} using the short-cadence (SC) photometric data of KIC 6382916 from {\it Kepler} telescope, obtained from BJD 2455064 to 2455091 (Quarter 02). We aimed to study the variations and interactions of these modes in a short timescale. Table \ref{tab:freq_solution} presents the frequencies and amplitudes of the 23 selected pulsation modes, as well as their absolute and relative variations ($\Delta A$, $\Delta A/\bar{A}$, $\Delta f$, and $\Delta f/\bar{f}$) and the Pearson correlation coefficients ($\rho_{A,f}$). We were able to extract variation information from the light curves for 19 of these pulsation modes. Further details can be found in the Supplementary Materials.

\begin{table*}[htp]
  \centering
  \caption{Overall Indicators of the 23 Pulsation Modes.}
  \label{tab:freq_solution}
  \begin{tabular}{l|ccccc|ccccr|r}
    \hline
    \hline
    ID   &Frequency&$\sigma_f$ &Amplitude&$\sigma_A$ & S/N &$\Delta f$ &$\Delta f/\bar{f}$ &$\Delta A$ &$\Delta A/\bar{A}$ &$\rho_{A,f}$ &Mark \\	
    {}   &($\cd$)&($\cd$)&(mmag)&(mmag)&{}&{($\cd$)}&{}&{(mmag)} &{} &{}\\ 
\hline 			     
    F0   & 4.9099	& 0.0004 & 81.2	 & 1.6	& 51.9  &0.013 &0.003 &4.094 &0.05 &-0.84 &$f_{0}$ \\
    F1   & 6.4320	& 0.0004 & 79.8	 & 1.4	& 58.2  &0.009 &0.001 &5.676 &0.07 &0.52 & $f_{1}$ \\
    F2   & 11.3418	& 0.0004 & 30.9	 & 0.6	& 55.5  &0.014 &0.001 &2.241 &0.07 &-0.03 & $f_{0}+f_{1}$ \\
    F3   & 1.5222	& 0.0004 & 16.3	 & 0.3	& 53.7  &0.029 &0.019 &4.359 &0.27 &0.97 & $-f_{0}+f_{1}$ \\
    F4   & 9.8197	& 0.0004 & 17.0	 & 0.3	& 53.7  &0.014 &0.001 &2.405 &0.14 &-0.67 & $2f_{0}$ \\
    F5   & 8.0374	& 0.0005 & 12.0	 & 0.3	& 42.1  &0.084 &0.01 &9.496 &0.75 &-0.92 & $f_{2}$ \\
    F6   & 12.8652	& 0.0004 & 13.3	 & 0.2	& 62.1  &0.039 &0.003 &5.090 &0.37 &0.97 & $2f_{1}$ \\
    F7   & 16.2519	& 0.0004 & 10.4	 & 0.2	& 55.9  &0.016 &0.001 &0.893 &0.09 &0.02 & $2f_{0}+f_{1}$ \\
    F8   & 7.9539	& 0.0004 &  7.0	 & 0.1	& 55.4  &0.092 &0.012 &4.618 &1.36 &0.80 & $-f_{0}+2f_{1}$ \\
    F9   & 3.3886	& 0.0005 &  5.1	 & 0.1	& 46.6  &0.072 &0.021 &3.347 &0.61 &-0.91 & $2f_{0}-f_{1}$ \\
    F10  & 17.7748	& 0.0004 &  5.36 & 0.09 & 62.8  &0.042 &0.002 &2.235 &0.41 &0.94 & $f_{0}+2f_{1}$ \\
    F11  & 12.9455	& 0.0005 &  3.33 & 0.07 & 47.0  &0.124 &0.01 &2.141 &1.25 &-0.87 & $f_{0}+f_{2}$ \\
    F12  & 22.6843	& 0.0004 &  4.94 & 0.09 & 58.4  &0.024 &0.001 &0.895 &0.18 &0.67 & $2f_{0}+2f_{1}$ \\
    F13  & 14.4704	& 0.0007 &  3.5  & 0.1	& 32.9  &0.100 &0.007 &2.742 &0.79 &-0.71 & $f_{1}+f_{2}$ \\
    F14$^{*}$  & 1.6044	& 0.0006 &  2.79 & 0.07 & 40.7 &--- &--- &--- &--- &--- & $-f_{1}+f_{2}$ \\
    F15$^{*}$  & 3.1222	& 0.0005 &  2.49 & 0.05 & 45.9 &--- &--- &--- &--- &--- & $-f_{0}+f_{2}$ \\
    F16$^{*}$  & 3.3063	& 0.0008 &  1.98 & 0.07 & 30.2 &--- &--- &--- &--- &--- & $f_{0}+f_{1}-f_{2}$ \\
    F17  & 14.7297	& 0.0005 &  2.38 & 0.06 & 42.5  &0.071 &0.005 &1.069 &0.47 &-0.50 & $3f_{0}$ \\
    F18  & 9.5582	& 0.0006 &  2.00 & 0.05 & 38.3  &0.118 &0.012 &0.949 &0.5 &-0.19 & $-f_{0}+f_{1}+f_{2}$ \\
    F19  & 19.2982	& 0.0005 &  2.69 & 0.06 & 48.7  &0.076 &0.004 &1.988 &0.67 &0.92 & $3f_{1}$ \\
    F20  & 24.2067	& 0.0004 &  3.04 & 0.06 & 52.4  &0.044 &0.002 &1.495 &0.48 &0.96 & $f_{0}+3f_{1}$ \\
    F21  & 21.1616	& 0.0005 &  1.97 & 0.04 & 48.4  &0.052 &0.002 &0.833 &0.41 &-0.77 & $3f_{0}+f_{1}$ \\
    F22$^{*}$  & 3.043	& 0.001	 &  1.32 & 0.06 & 22.7 &--- &--- &--- &--- &--- & $-2f_{0}+2f_{1}$ \\
\hline
  \end{tabular}
\\
\footnotesize{Note: $^{*}$ denotes the pulsation modes which we cannot extract their amplitude and frequency variations because of the strong interactions and low SNRs.}
\end{table*}

\section{Results and Discussions}

\subsection{Resonating Integration Mode}

Based on the proposed framework of the Resonating Integration (RI) mode by \citet{Niu2022}, two important behaviors were predicted on short timescales for the pulsation modes F5 ($f_{2}$) and F8 ($-f_{0} + 2f_{1}$): (i) as the non-radial part of a RI mode, F5 should exhibit amplitude and frequency modulations; (ii) as the superposition and resonance of the radial part of a RI mode and the combination of $f_{0}$ and $f_{1}$, F8 should have a dual identity and exhibit frequency splitting while also being modulated (if two parts of it are out of sync).

\begin{figure*}
  \gridline{\fig{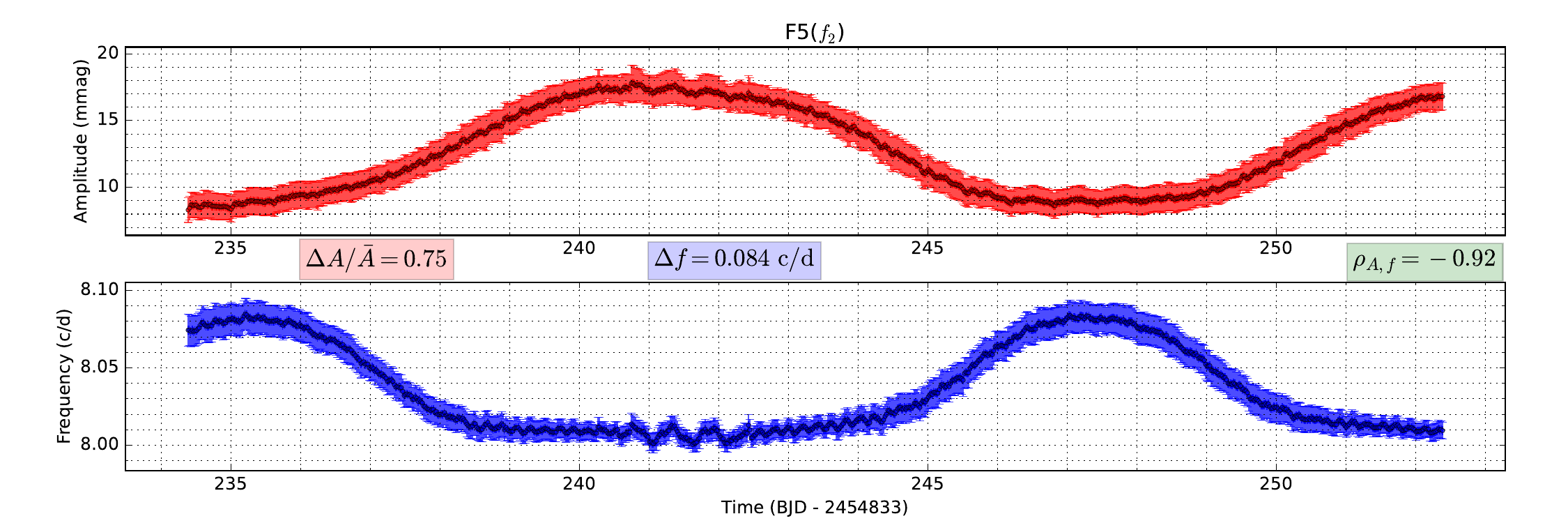}{0.8\textwidth}{(a)}}
  \gridline{\fig{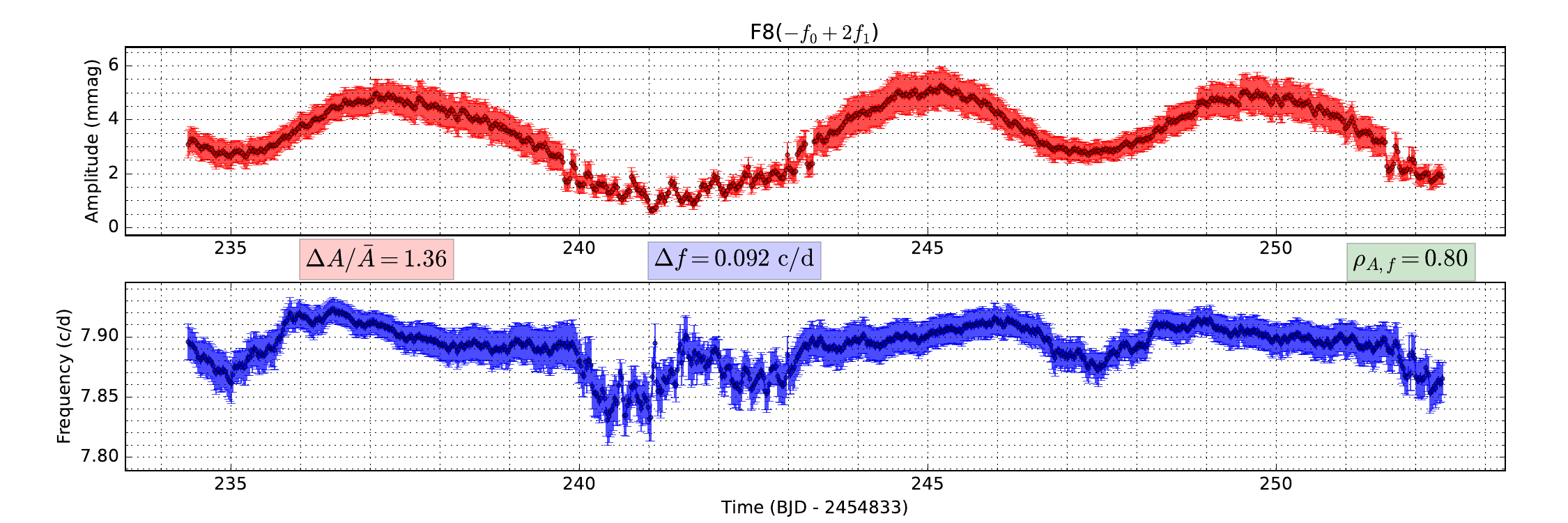}{0.8\textwidth}{(b)}}
  \gridline{\fig{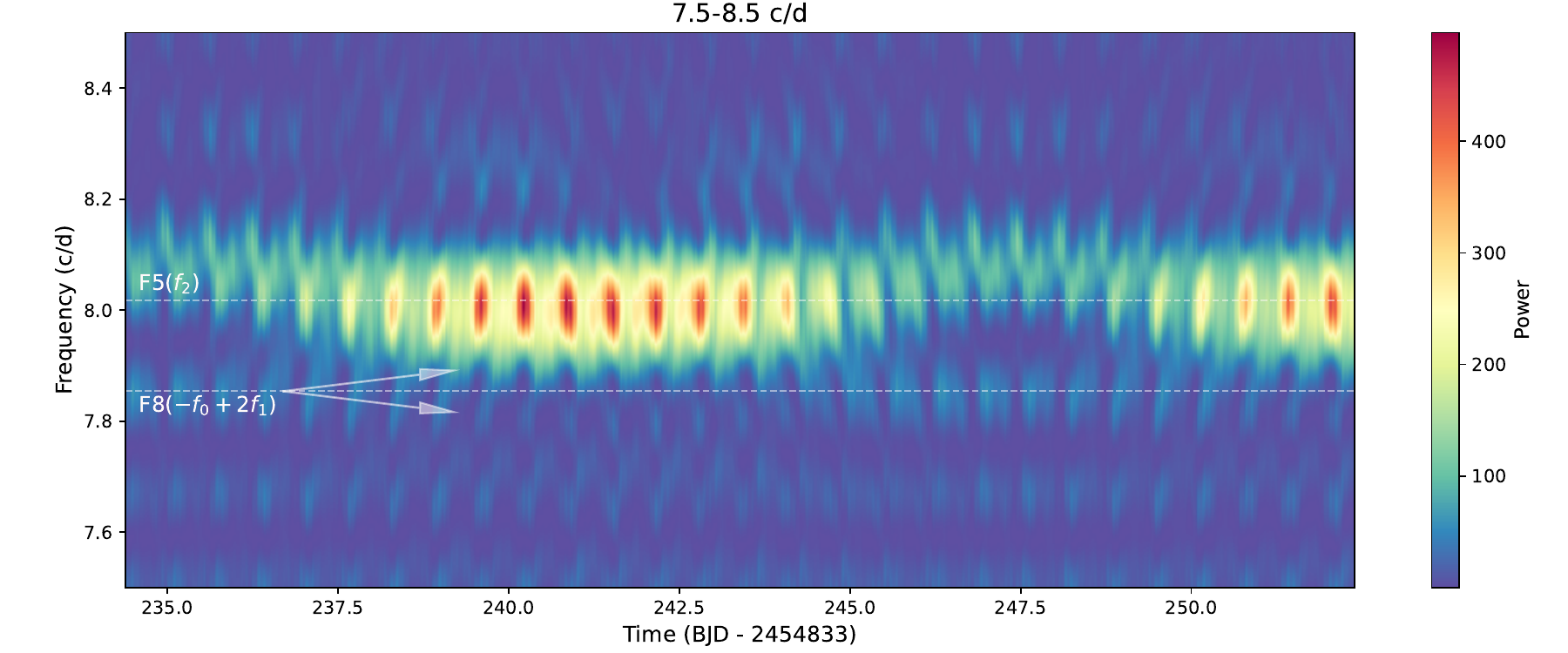}{0.8\textwidth}{(c)}}
  \caption{Variation of the amplitudes and frequencies of F5 and its partner F8. Subfigure (a) represents the amplitude and frequency variations of F5; subfigure (b) represents the amplitude and frequency variations of F8; subfigure (c) is the time-frequency diagram which presents interaction details in $7.5-8.5\ \cd$.}
  \label{fig:var_f2}
  \end{figure*}

Figure \ref{fig:var_f2} presents the amplitude and frequency variations, as well as the interaction details of F5 and F8.
Subfigure (a) and (b) show that F5 and F8 exhibit distinct amplitude and frequency modulations as predicted.
Subfigure (c) reveals that F8 splits into two parts after approximately 235 days. The first part, which is opposite to the modulation phase of F5, approaches and resonates with F5. The second part, which is the same as the modulation phase of F5, seems to approach and resonate with another pulsation mode. After about 246 days, these two parts of F8 merge, and the amplitude of F5 remains low due to decoupling from the first part of F8.
However, for F8, its amplitude is not completely anti-correlated with F5, as evident from the pit between 246 and 249 days. This indicates that the first part of F8 has its own modulation rhythm, independent of F5, which arises from $f_{0}$ and $f_{1}$.

Furthermore, in subfigures (a) and (b), the amplitude and frequency fluctuations of F5 (around 241-242 days) and F8 (around 240-243 days) are caused by strong resonant interactions with adjacent pulsation modes, making it difficult to accurately distinguish them in these time regions.\footnote{The amplitude and frequency variations of F14, F15, F16, and F22 could not be extracted for the same reason. In the lower frequency region, there are more pulsation modes and complex interactions (see in Appendix, Figure \ref{fig:var_2-3}).}

In Figures \ref{fig:var_amp_freq01}, \ref{fig:var_amp_freq02}, \ref{fig:var_amp_freq03}, and \ref{fig:var_amp_freq04}, it can be observed that the amplitudes and frequencies of many pulsation modes exhibit strong (anti)correlation. Specifically, the pulsation modes F3, F5, F6, F9, F10, F19, and F20 show significant (anti)correlations, with their $\vert \rho_{A,f} \vert$ values exceeding 9.0. The amplitude and frequency (anti)correlations of these modes can be consistently explained by the framework provided by F5 and F8. The intrinsic frequency modulation of these modes leads to resonances between these modes and their partners. These resonances result in the growth of amplitudes. Although the modulation in frequency appears to be of first nature, it presents another problem that needs to be addressed, which comes from the non-correlation of the frequencies and their harmonics.

\subsection{Harmonics and Combinations}
In pulsating stars, such as Cepheids \citep{Rathour2021}, RR Lyrae stars \citep{Kurtz2016}, $\delta$ Scuti stars \citep{Breger2014}, high-amplitude $\delta$ Scuti stars \citep{Niu2017}, SX Phoenicis stars \citep{Xue2020}, $\gamma$ Dor stars \citep{Kurtz2015}, pulsating white dwarfs \citep{Wu2001}, $\beta$ Cep stars \citep{Degroote2009}, and slowly pulsating B stars \citep{Papics2017}, harmonics of independent pulsation modes are quite common. These harmonics arise from the nonsinusoidal nature of light curves and indicate the nonlinearity of the star's pulsation. Therefore, harmonics are not considered as intrinsic stellar pulsation modes \citep{Brickhill1992, Wu2001} and they should mimic the behaviors of their parent pulsation modes. However, recent research has shown that as the order of the harmonics increases, the amplitude and frequency of the harmonics exhibit uncorrelated variations compared to their parent pulsation mode \citep{Niu2023}.

\begin{figure*}[htp]
  \centering
  \includegraphics[width=0.8\textwidth]{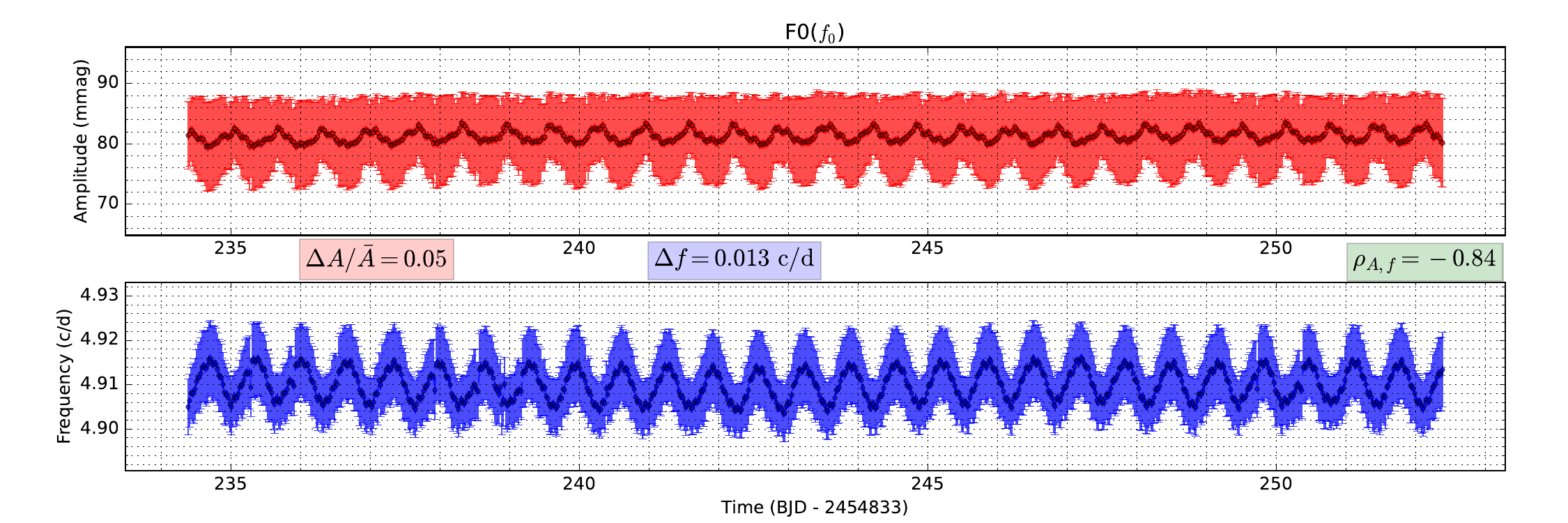}
  \includegraphics[width=0.8\textwidth]{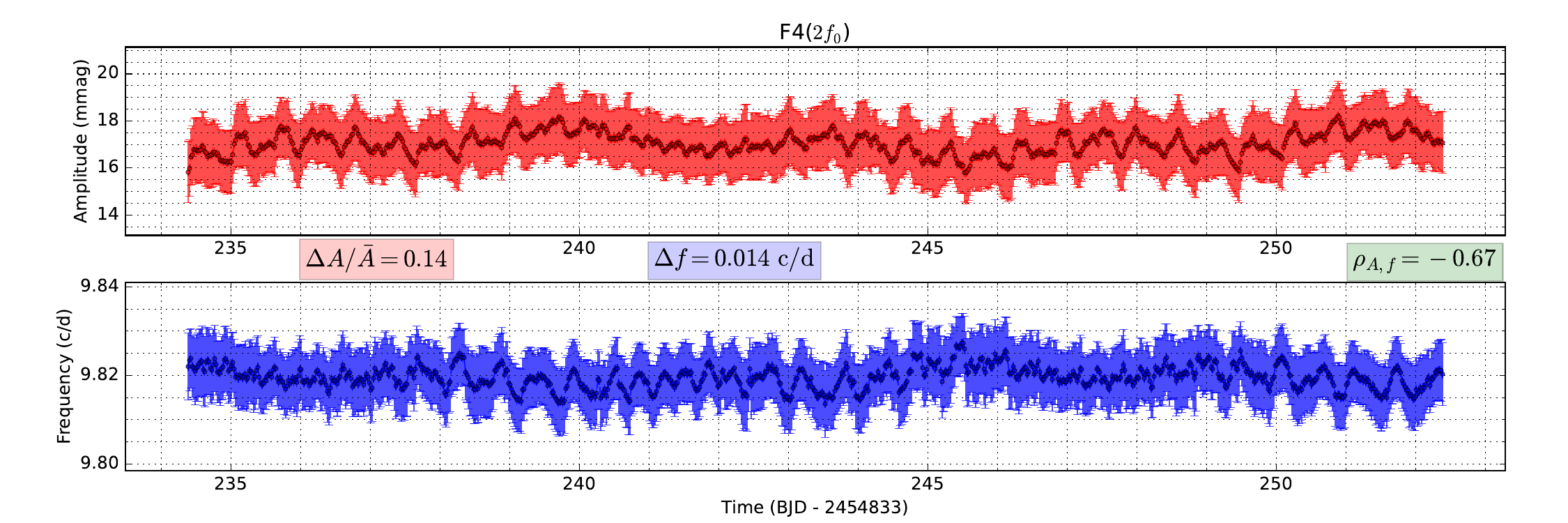}
  \includegraphics[width=0.8\textwidth]{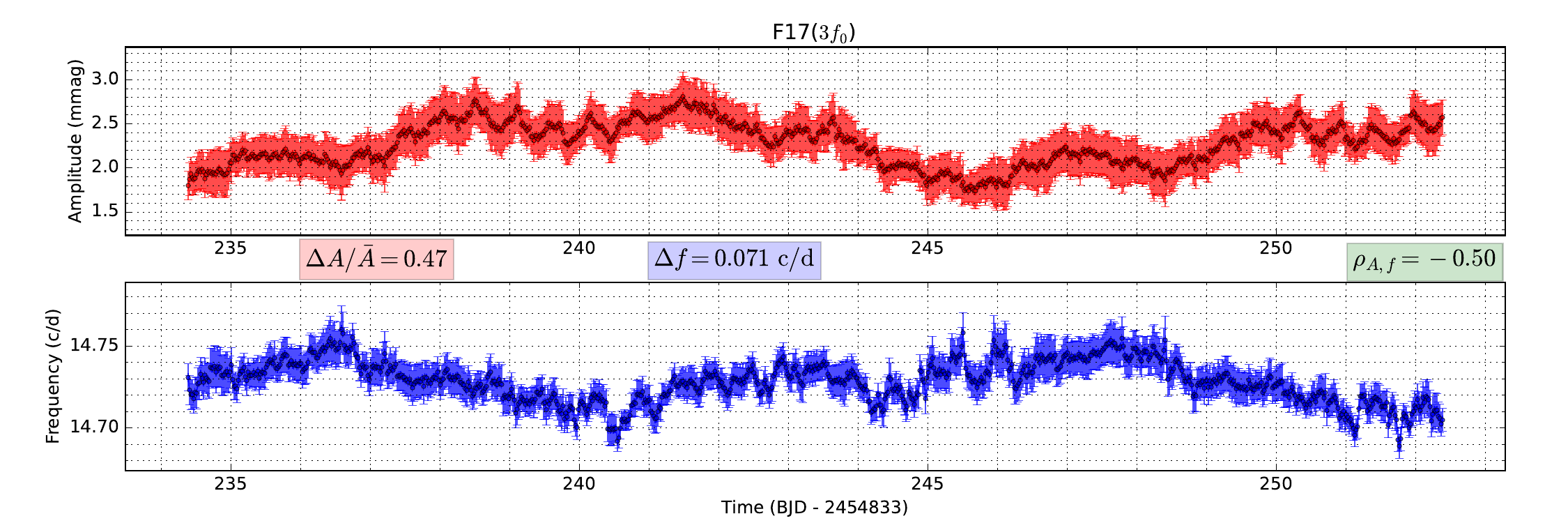}
  \caption{Amplitude and frequency variations of the independent pulsation mode $f_0$ and its first and second harmonics ($2f_0$ and $3f_0$).}
  \label{fig:var_f0}
\end{figure*}

\begin{figure*}[htp]
  \centering
  \includegraphics[width=0.8\textwidth]{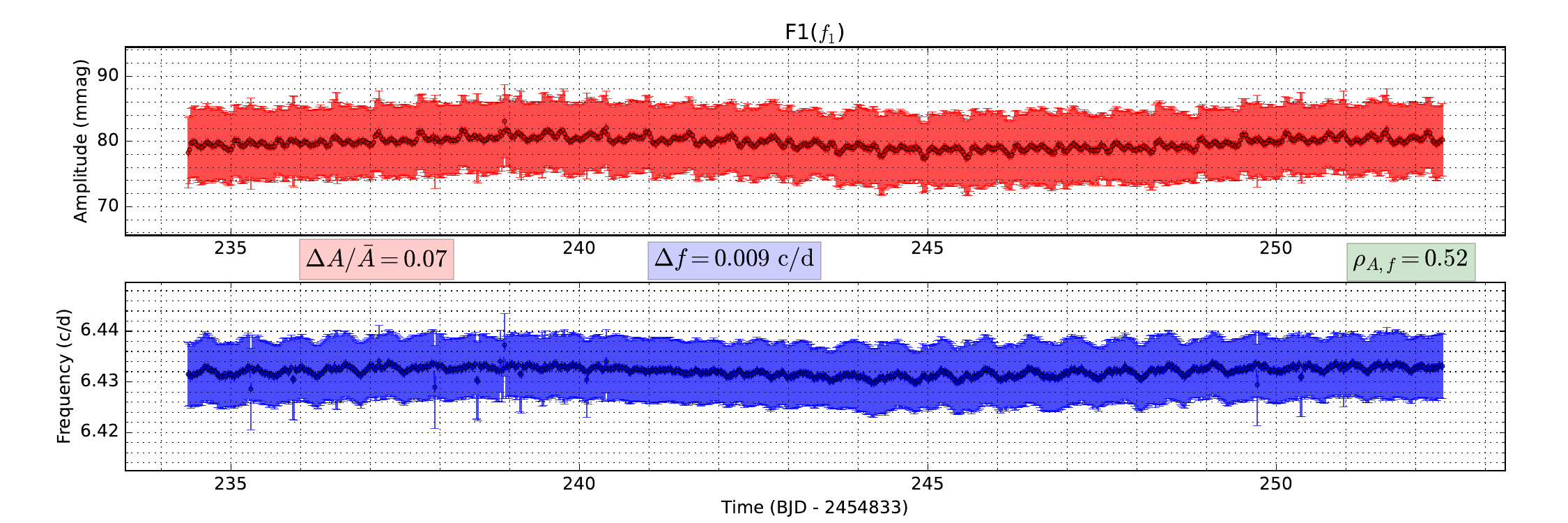}
  \includegraphics[width=0.8\textwidth]{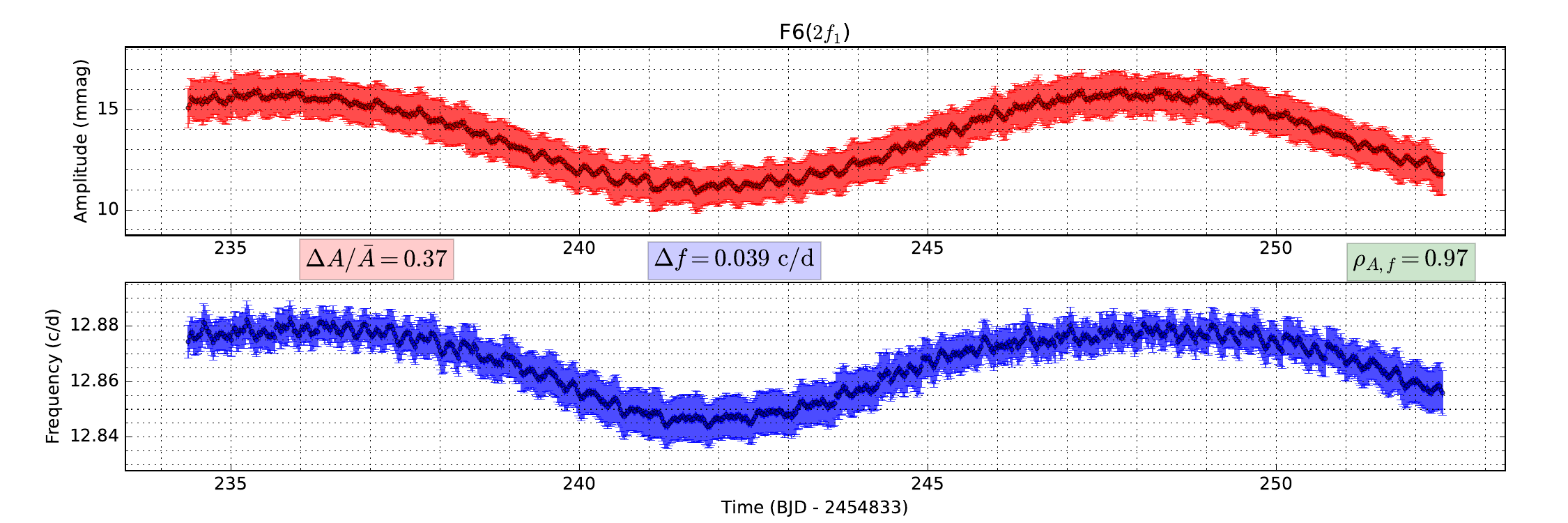}
  \includegraphics[width=0.8\textwidth]{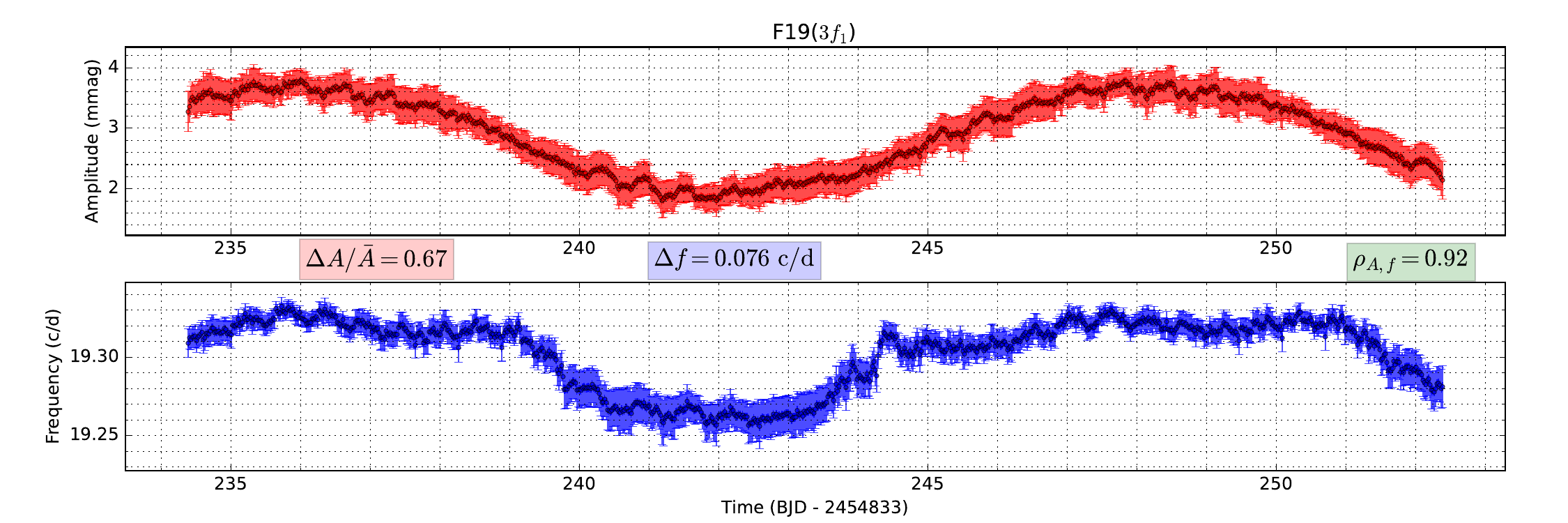}
  \caption{Amplitude and frequency variations of the independent pulsation mode $f_1$ and its first and second harmonics ($2f_1$ and $3f_1$).}
  \label{fig:var_f1}
\end{figure*}

For KIC 6382916, the amplitude and frequency variations of the fundamental and first overtone pulsation modes ($f_0$ and $ f_1$) along with their first and second harmonics ($2f_0$, $3f_0$, $2f_1$, and $3f_1$) are presented in Figures \ref{fig:var_f0} and \ref{fig:var_f1}. It is evident that the harmonics of $f_0$ and $f_1$ show uncorrelated amplitude and frequency variations compared to their parent modes, even at the first and second harmonics. This challenges the common understanding derived from stellar pulsation theory, thus broadening our perception of harmonics.

Furthermore, the amplitude and frequency variations of $2f_1$ and $3f_1$ exhibit a high degree of correlation, indicating that $2f_1$ behaves like an independent parent pulsation mode of $3f_1$. In general, the pulsation mode $3f_1$ can be represented by $3 \cdot (f_1)$ and $(2f_1)+(f_1)$, which are equivalent representations in common. However, in this case, the correct representation of $3f_1$ is $(2f_1)+(f_1)$. This challenges the usual frequency identification of harmonics in the pre-whitening process and subsequently the identification of independent pulsation modes.
Moreover, if we adhere to the principle that frequency modulation is of first nature and explain amplitude modulations through resonances, the physical origin of frequency modulation in $2f_1$ becomes a question. It should be noted that $f_1$ does not exhibit obvious frequency modulation like $2f_1$.

The aforementioned case is not unique. It is observed that the pulsation modes F10 ($f_0+2f_1$) and F20 ($f_0+3f_1$) also exhibit highly correlated amplitude and frequency variations. This suggests that $f_0+3f_1$ can be represented as the sum of $(f_0+2f_1)$ and $(f_1)$, and $f_0+2f_1$ should be considered as an independent parent pulsation mode of $f_0+3f_1$.

\subsection{Overview of the Interactions}
\begin{figure*}[htp]
  \centering
  \includegraphics[width=0.6\textwidth]{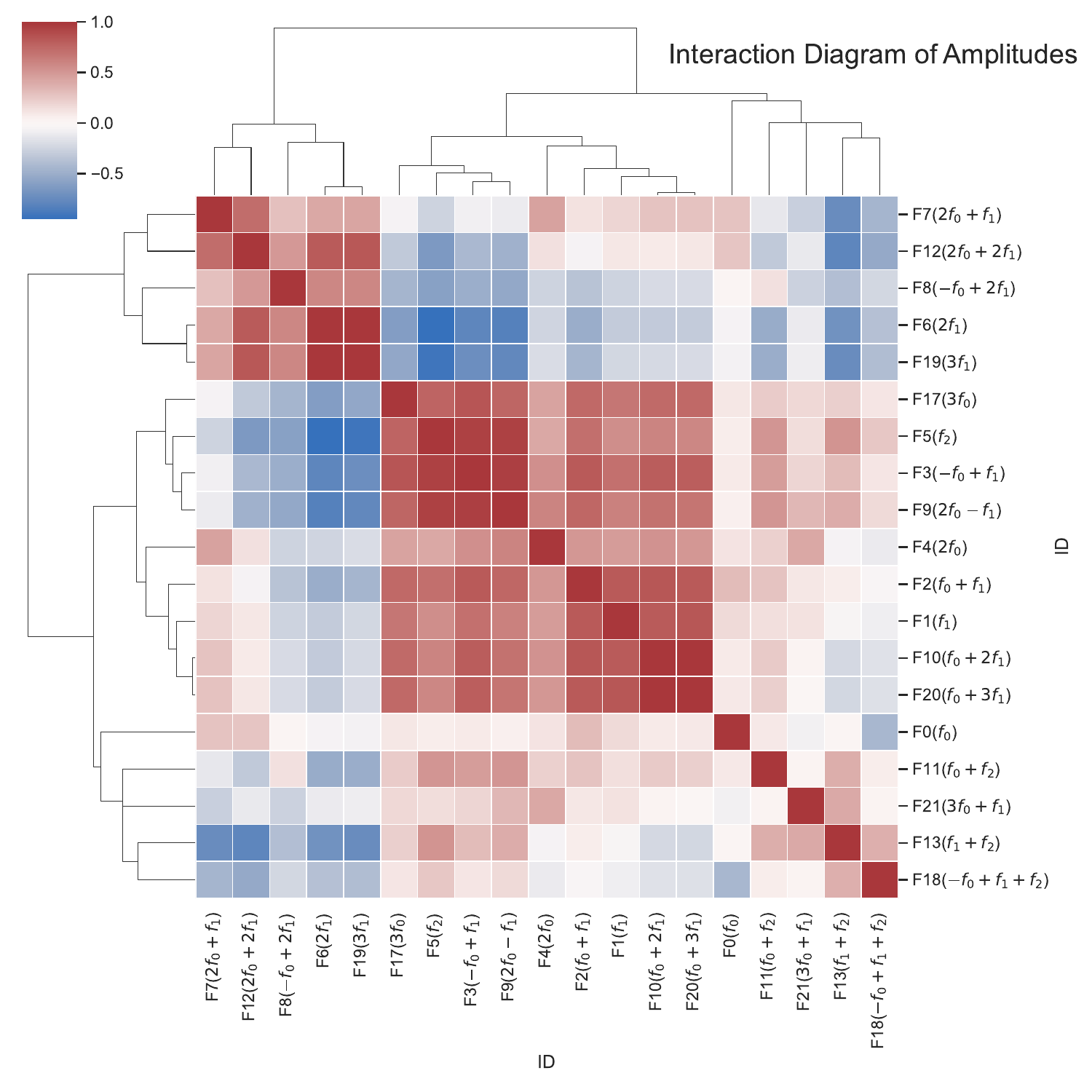}
  \includegraphics[width=0.6\textwidth]{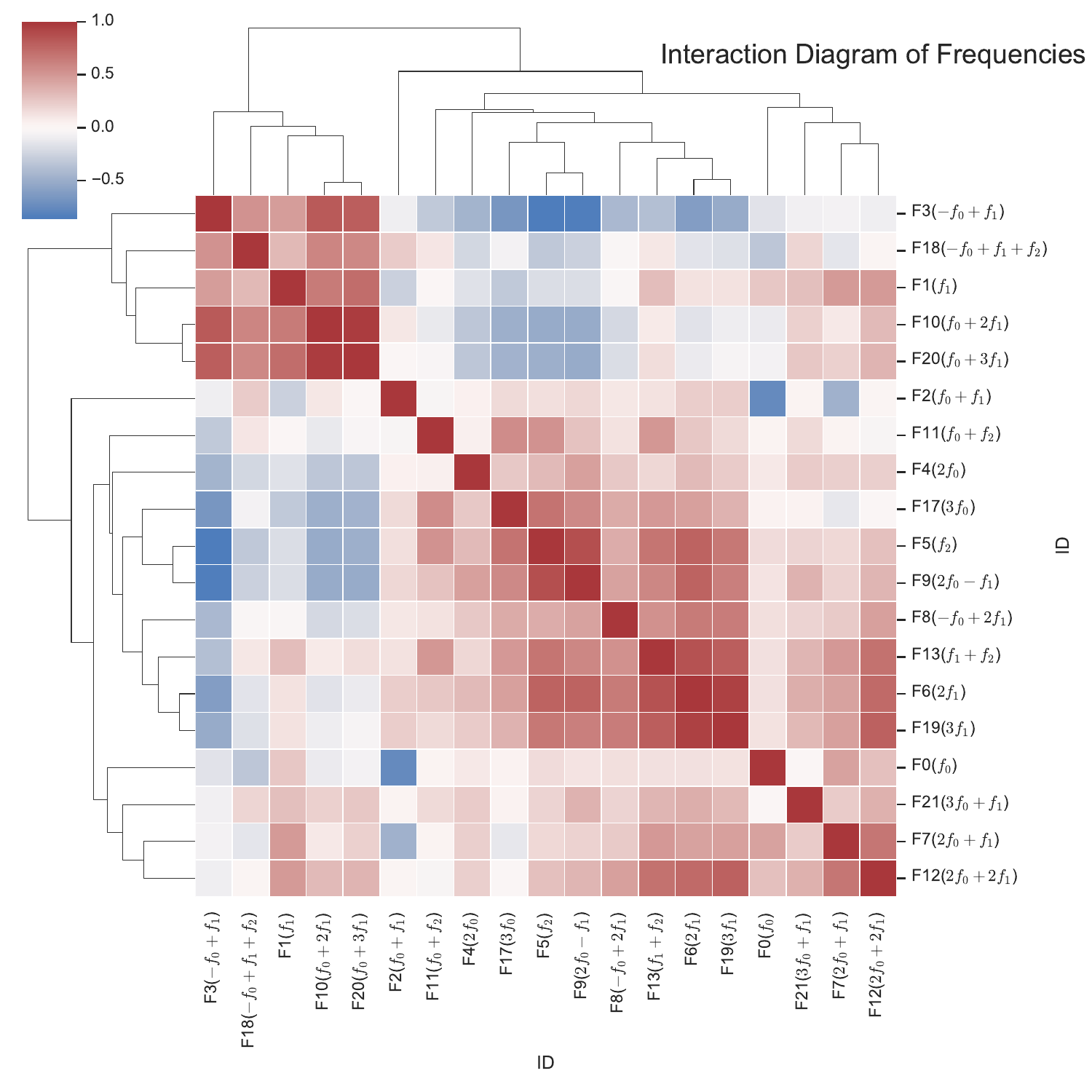}
  \caption{Interaction diagrams of amplitudes and frequencies of the 19 pulsation modes.}
  \label{fig:ID_amp_freq}
\end{figure*}

The interactions between the pulsation modes can be clearly represented in the amplitude and frequency interaction diagrams of the 19 pulsation modes shown in Figure \ref{fig:ID_amp_freq}. These diagrams cluster together pulsation modes with similar amplitude and frequency variation patterns. It is worth noting that these interactions are complex, indicating that the harmonics and combinations are more intricate than commonly assumed, and they contain rich information about the nonlinear characteristics. Furthermore, the resulting interaction relationships differ from those observed in longer timescales.
Within the interaction diagrams, only F6/F19 and F10/F20 exhibit the most similar behaviors in terms of both amplitude and frequency variations. This observation provides valuable insights into reevaluating the physical nature of the harmonics and combinations, as discussed earlier.

In the interaction diagram of amplitudes, although the modes related to $f_2$ (F5, F11, F13, and F18) are not clustered together as observed previously in \citet{Niu2022}, they exhibit anti-correlated behavior with the modes in the first cluster (F7, F12, F8, F6, and F19). This suggests the presence of energy transfer between the $f_2$-related and some $f_1$-related modes. Specifically, if we consider F7 ($2f_0 + f_1$) and F12 ($2f_0 + 2f_1$) as $(2f_0 - f_1) + (2f_1)$ and $(2f_0 - f_1) + (3f_1)$, respectively\footnote{This decomposition is inspired by the relationship between $2f_1$ and $3f_1$.}, we can observe that the modes in the first cluster are all related to $2f_1$ and $3f_1$. Based on the relation $3f_1 \equiv (2f_1) + (f_1)$, it is evident that the most significant energy transfer channel of the first cluster is $2f_1$ rather than $f_1$.

In the interaction diagram of frequencies, it can be observed that F3 and F2 exhibit the strongest interactions with other pulsation modes. F3 ($-f_0 + f_1$) has the largest number of anti-correlated pulsation modes. On the other hand, F2 ($f_0 + f_1$) is anti-correlated to F0 ($f_0$), F1 ($f_1$), and F7 ($2f_0 + f_1$), which is quite strange if we note that the first two are the parent modes of F2. This finding challenges our traditional understanding of combination modes in pulsating stars, suggesting the presence of an undiscovered mechanism when these modes combine.

The frequency anti-correlation between F0 ($f_0$) and F2 ($f_0 + f_1$) is particularly noteworthy. Almost all pulsation modes are affected by a small frequency modulation of approximately 0.004 $\cd$ as evident in F0 and F2. This modulation exhibits a period of 0.6570 day and appears to be the background in the time-frequency diagram (see, e.g., Figures \ref{fig:var_2-3}). The physical origin of this modulation is still unknown.

\subsection{Summary}
In summary, a comprehensive analysis of the pulsation modes in KIC 6382916 reveals a striking analogy to the surface of the sea. On larger timescales, the pulsation modes exhibit a seemingly calm or gradual trend. However, when observed on smaller timescales, they undergo violent fluctuations, unveiling intricate pulsation mode interactions. These interactions manifest as amplitude and frequency modulations, presenting a dual identity of certain modes and posing challenges to conventional concepts like harmonics and combinations. Despite the complexity inherent in these phenomena, discernible patterns emerge, offering valuable guidance for future research in nonlinear asteroseismology.
This profound understanding of the pulsation modes in KIC 6382916 provides a foundation for unraveling the mysteries of stellar pulsation and opens up exciting avenues for further exploration in the field.

\section*{Acknowledgments}
J.S.N. acknowledges support from the National Natural Science Foundation of China (NSFC) (No. 12005124 and No. 12147215). H.F.X. acknowledges support from the Scientific and Technological Innovation Programs of Higher Education Institutions in Shanxi (STIP) (No. 2020L0528) and the Applied Basic Research Programs of Natural Science Foundation of Shanxi Province (No. 202103021223320).
The authors acknowledge the Kepler Science team and everyone who has contributed to making the Kepler mission possible. Funding for the Kepler mission is provided by NASA’s Science Mission Directorate. 

\software{{\tt astropy} \citep{astropy}, {\tt seaborn} \citep{seaborn}, {\tt Lightkurve} \citep{lightkurve}, {\tt NumPy} \citep{numpy}, {\tt SciPy} \citep{scipy}, {\tt matplotlib} \citep{matplotlib}}

\clearpage

\appendix
\setcounter{figure}{0}
\setcounter{table}{0}
\renewcommand{\thefigure}{A\arabic{figure}}
\renewcommand{\thetable}{A\arabic{table}}

\section{Methods}

\subsection{Photometric Data Reduction}
\begin{figure*}[htp]
  \centering
  \includegraphics[width=0.9\textwidth]{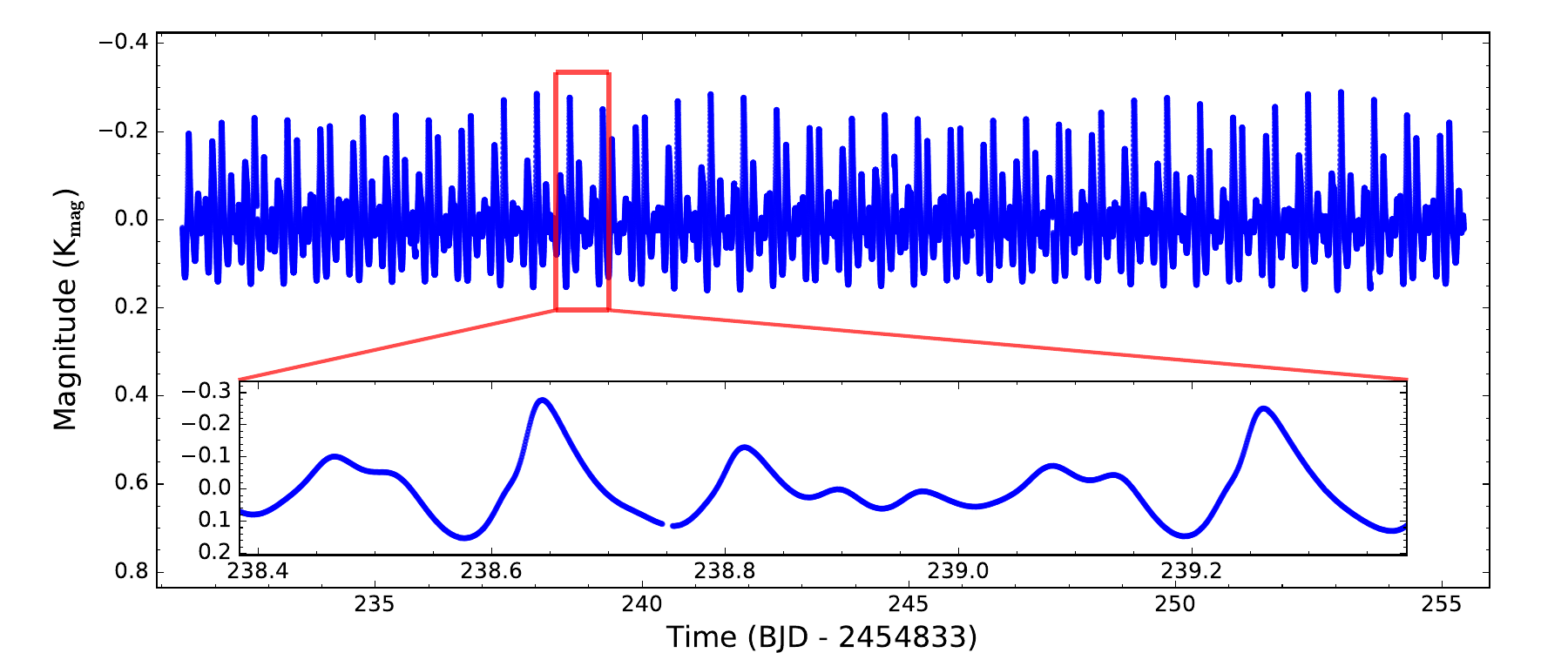}
  \includegraphics[width=0.9\textwidth]{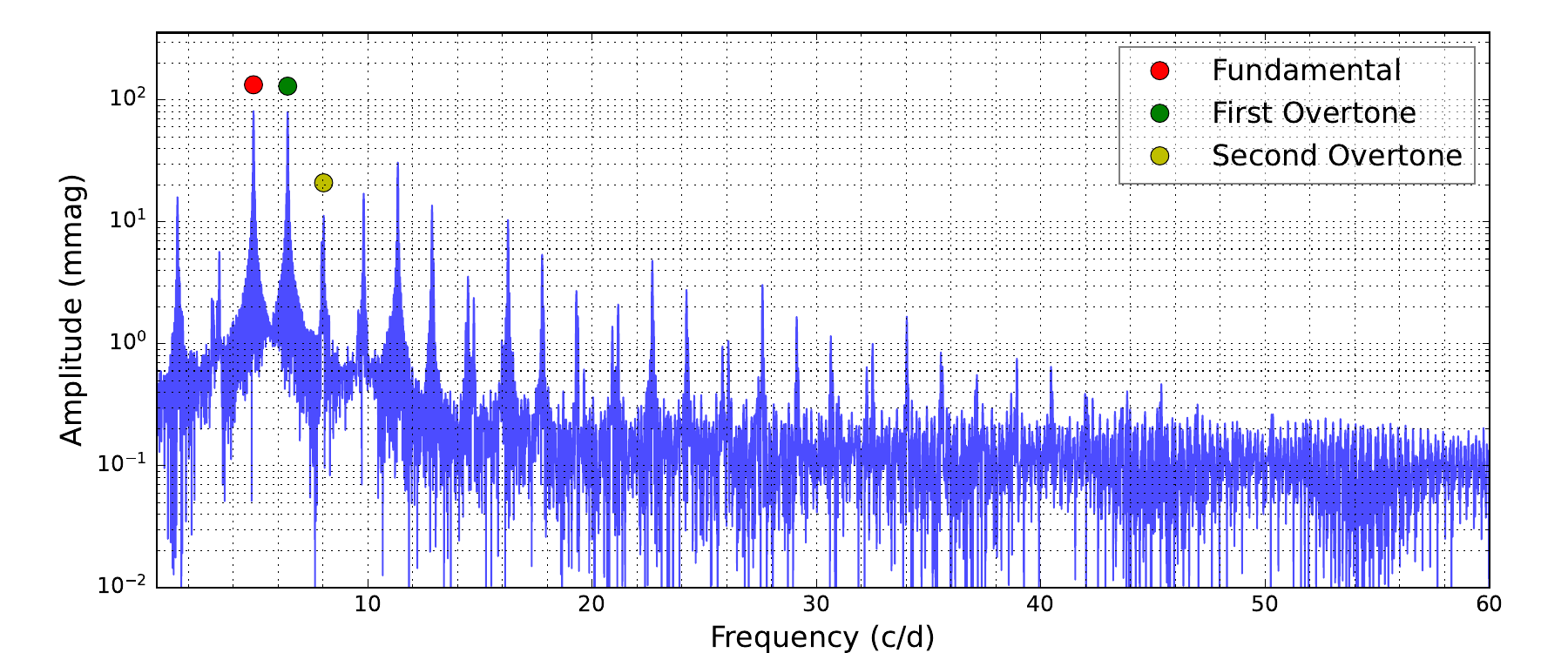}
  \caption{Overview of the normalized SC data in time domain and frequency domain of KIC 6382916. The region surrounded by the red rectangular boxes is selected to zoom in the details of the observed light curves. The 3 independent frequencies are marked as colored dots.}
  \label{fig:overview}
\end{figure*}

This study utilized short-cadence (SC) photometric data of KIC 6382916 obtained from the {\it Kepler space telescope} \citep{Kepler01,Kepler02}. The data span from BJD 2455064 to 2455091 (Quarter 02, with a small portion of discontinuous data removed, approximately 2 days in length) and were downloaded from the Mikulski Archive for Space Telescope (MAST)\footnote{\url{http://archive.stsci.edu/kepler}}. The data, in the form of reduced BJD and magnitudes, were then normalized to have a zero mean. Figure~\ref{fig:overview} provides an overview of the normalized SC data in both the time and frequency domains.

All the aforementioned normalized SC data were pre-whitened to extract the frequencies and amplitudes of the pulsation modes. The process was terminated once all the frequencies studied in \citet{Niu2022} (a total of 23 frequencies) had been identified (see Table \ref{tab:freq_solution}). For ease of comparison, the IDs of the pulsation modes in this work were arranged in the same order as in \citet{Niu2022}, rather than in descending order of amplitudes based on the SC data.

The Short-Time Fourier Transformation \citep{Bowman2016,Zong2018,Niu2022,Niu2023} was then applied to the normalized SC data to determine the variations of the amplitudes and frequencies. In this process, a time window of 6 days was moved from the start to the end time of the SC data, with a step of 0.02 days. At each step, the pre-whitening process was conducted to extract the amplitudes and frequencies of the specific 23 pulsation modes. The phase was considered as a free parameter and was not the focus of this study.

Next, we collected the amplitudes and frequencies from each moving window for each of the 19 pulsation modes in 23. The times were defined as the midpoints of the windows. The remaining 4 pulsation modes (F14, F15, F16, and F22) exhibited complex interactions with the nearby pulsation modes, making it difficult to identify and distinguish them from each other (see, e.g., Figure \ref{fig:var_2-3}).
The variations in the amplitudes and frequencies of the 19 pulsation modes are shown in Figures \ref{fig:var_amp_freq01}, \ref{fig:var_amp_freq02}, \ref{fig:var_amp_freq03}, and \ref{fig:var_amp_freq04}.

In this study, the pre-whitening process was conducted using Fourier decomposition, which can be represented by the formula:
\begin{equation}
  \label{eq:Fourier_de}
  m = m_{0} + \sum A_{i} \sin \left[ 2 \pi (f_{i} t + \phi_{i}) \right] ~,
\end{equation}
where $m_0$ is the shifted value, $A_{i}$ is the amplitude, $f_{i}$ is the frequency, and $\phi_{i}$ is the corresponding phase.
The uncertainties of the amplitudes ($\sigma_A$), frequencies ($\sigma_f$), and phases ($\sigma_{\phi}$) throughout this study were estimated using the framework introduced in \citet{Niu2022} and \citet{Niu2023}.

\subsection{Overall Indicators of Pulsation Modes}

\begin{figure*}[htp]
  \centering
  \includegraphics[width=0.8\textwidth]{fig/amp_freq_absvar_0.pdf}
  \includegraphics[width=0.8\textwidth]{fig/amp_freq_absvar_1.pdf}
  \includegraphics[width=0.8\textwidth]{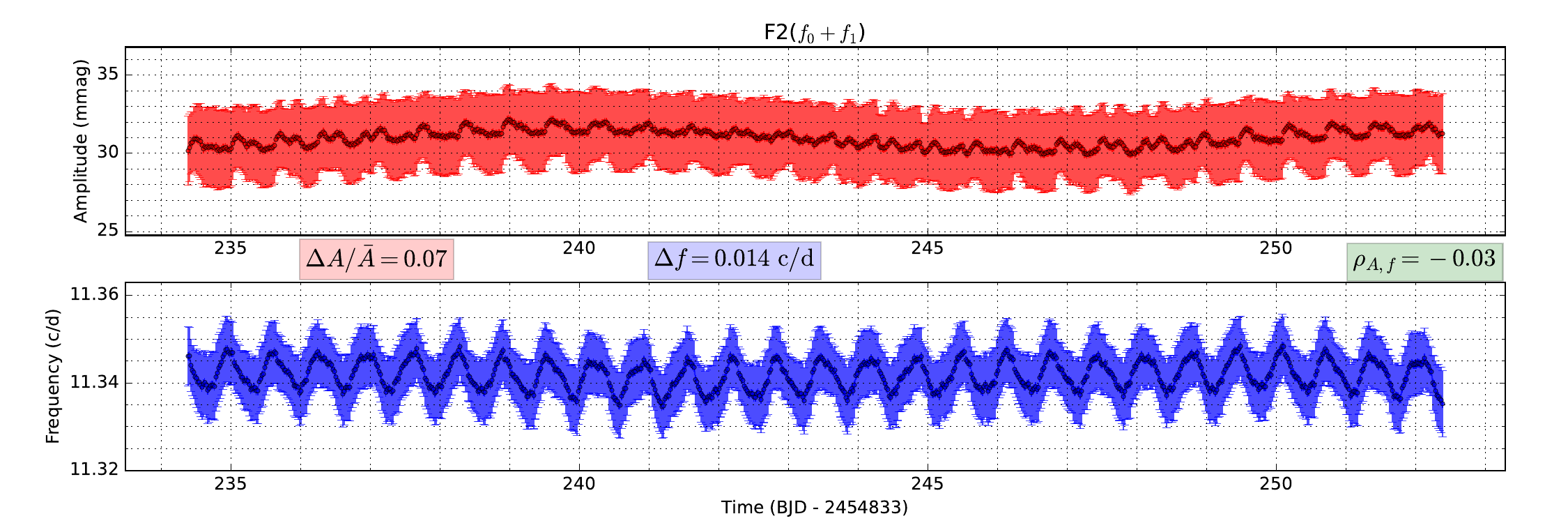}
  \includegraphics[width=0.8\textwidth]{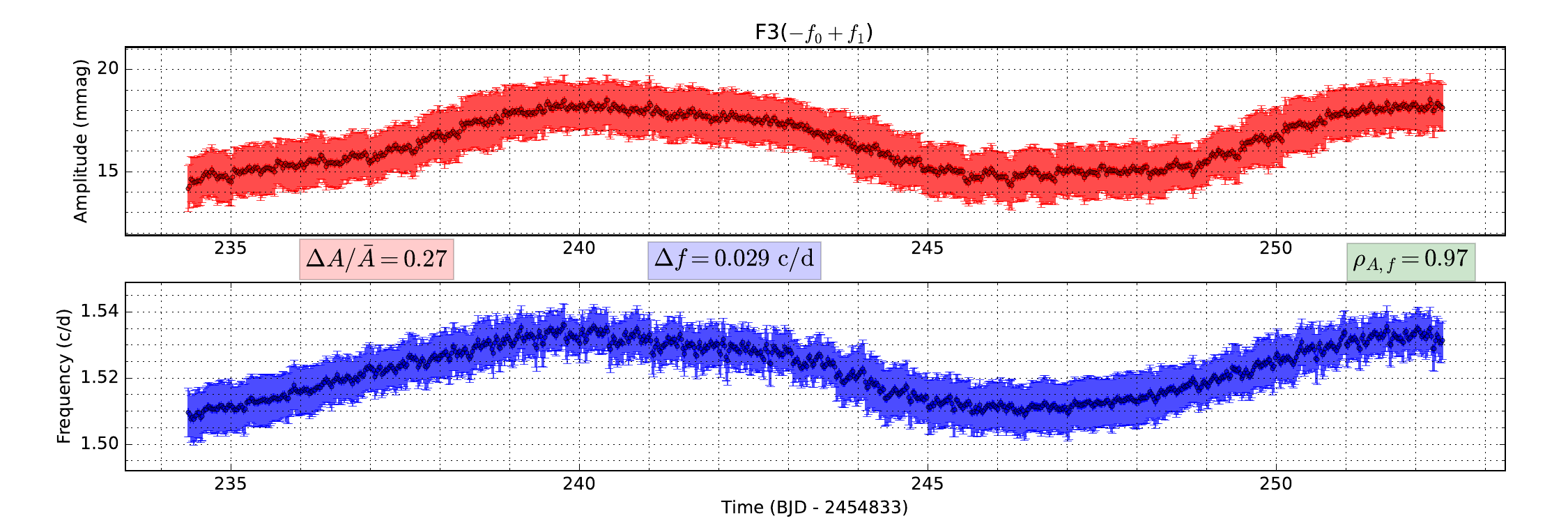}
  \includegraphics[width=0.8\textwidth]{fig/amp_freq_absvar_4.pdf}
  \caption{Amplitude and frequency variations of the 19 pulsation modes, Part I.}
  \label{fig:var_amp_freq01}
\end{figure*}

\begin{figure*}[htp]
  \centering
  \includegraphics[width=0.8\textwidth]{fig/amp_freq_absvar_5.pdf}
  \includegraphics[width=0.8\textwidth]{fig/amp_freq_absvar_6.pdf}
  \includegraphics[width=0.8\textwidth]{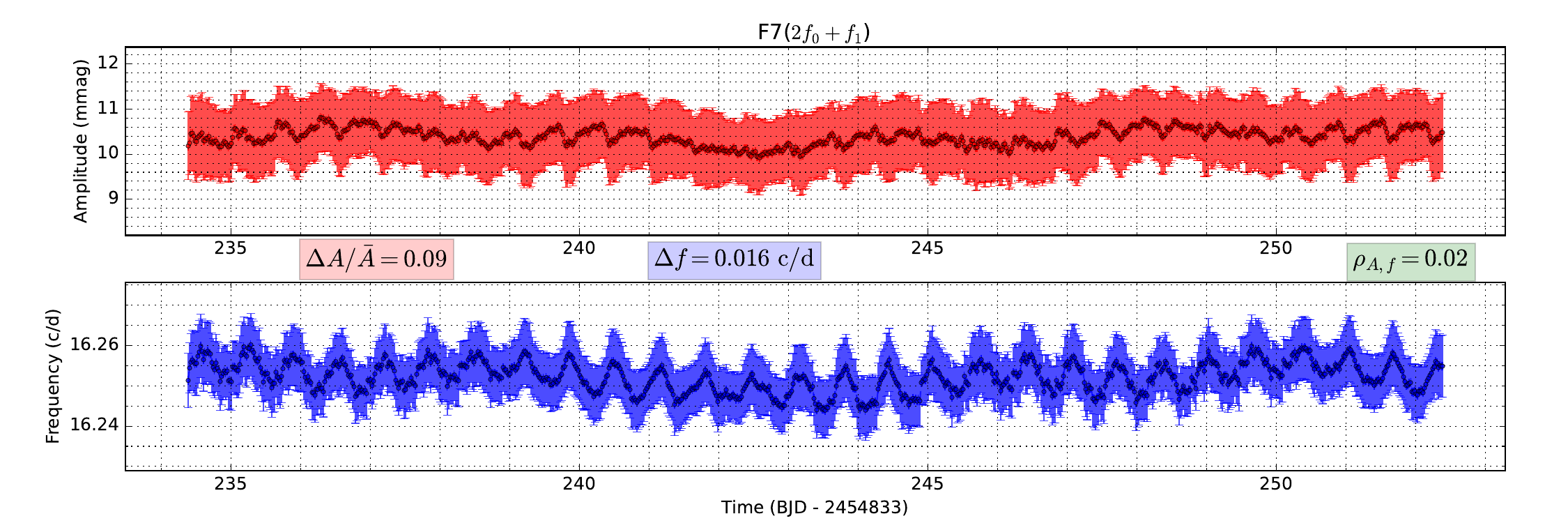}
  \includegraphics[width=0.8\textwidth]{fig/amp_freq_absvar_8.pdf}
  \includegraphics[width=0.8\textwidth]{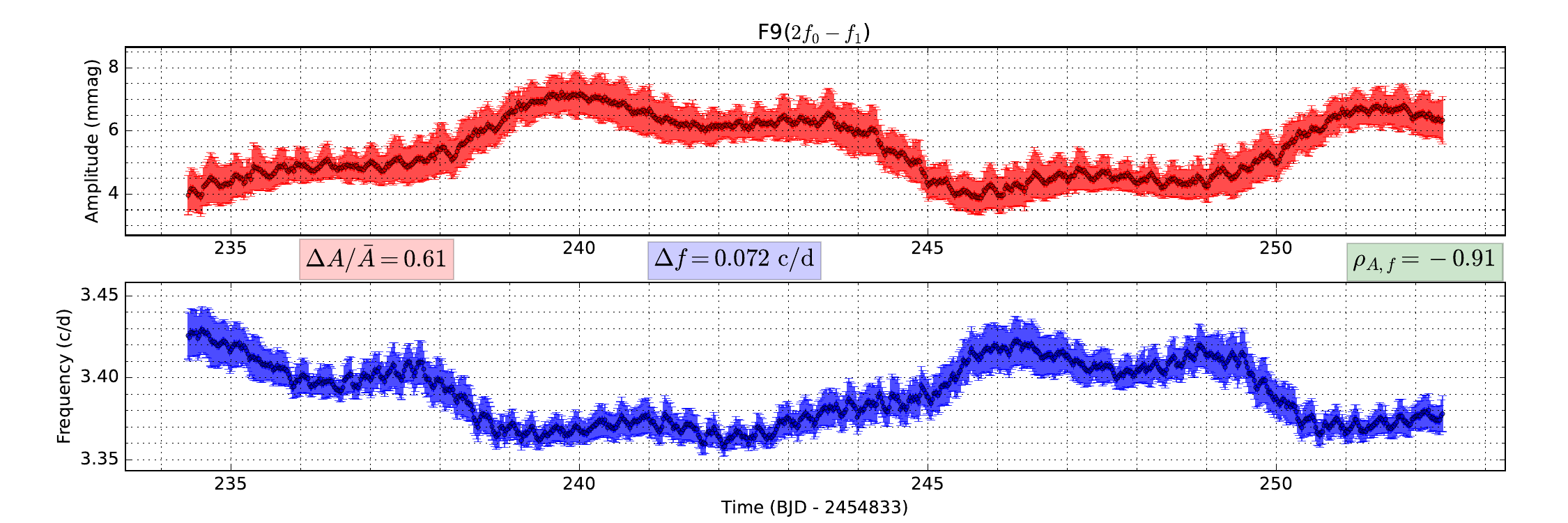}
  \caption{Amplitude and frequency variations of the 19 pulsation modes, Part II.}
  \label{fig:var_amp_freq02}
\end{figure*}

\begin{figure*}[htp]
  \centering
  \includegraphics[width=0.8\textwidth]{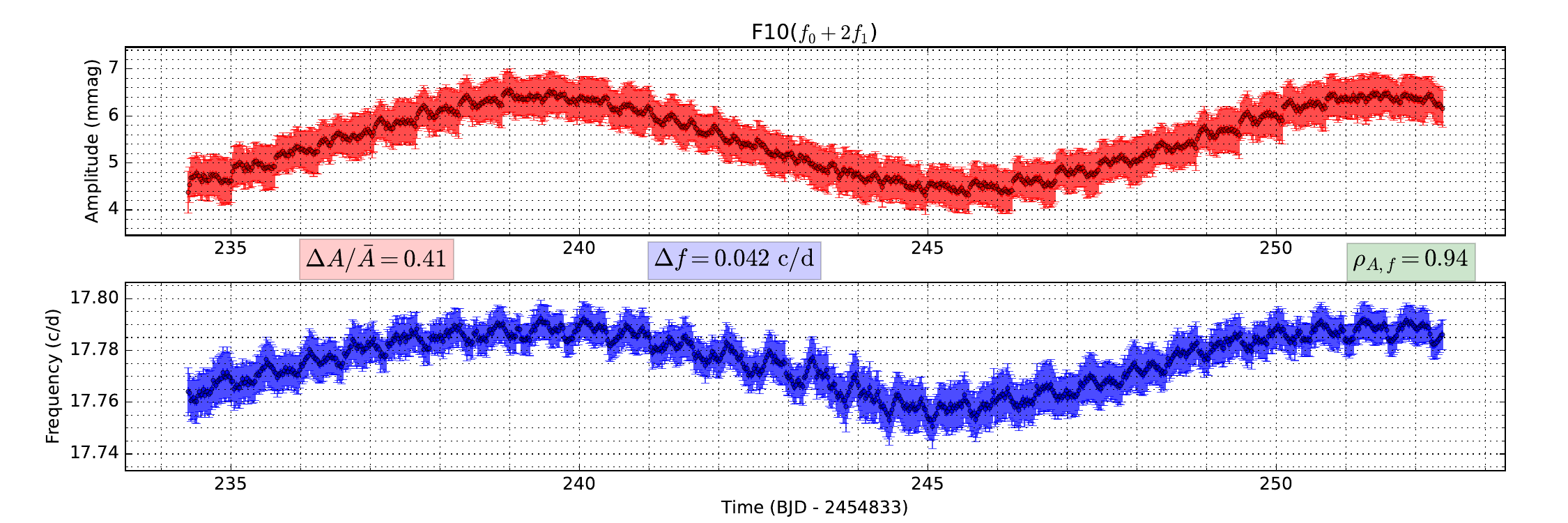}
  \includegraphics[width=0.8\textwidth]{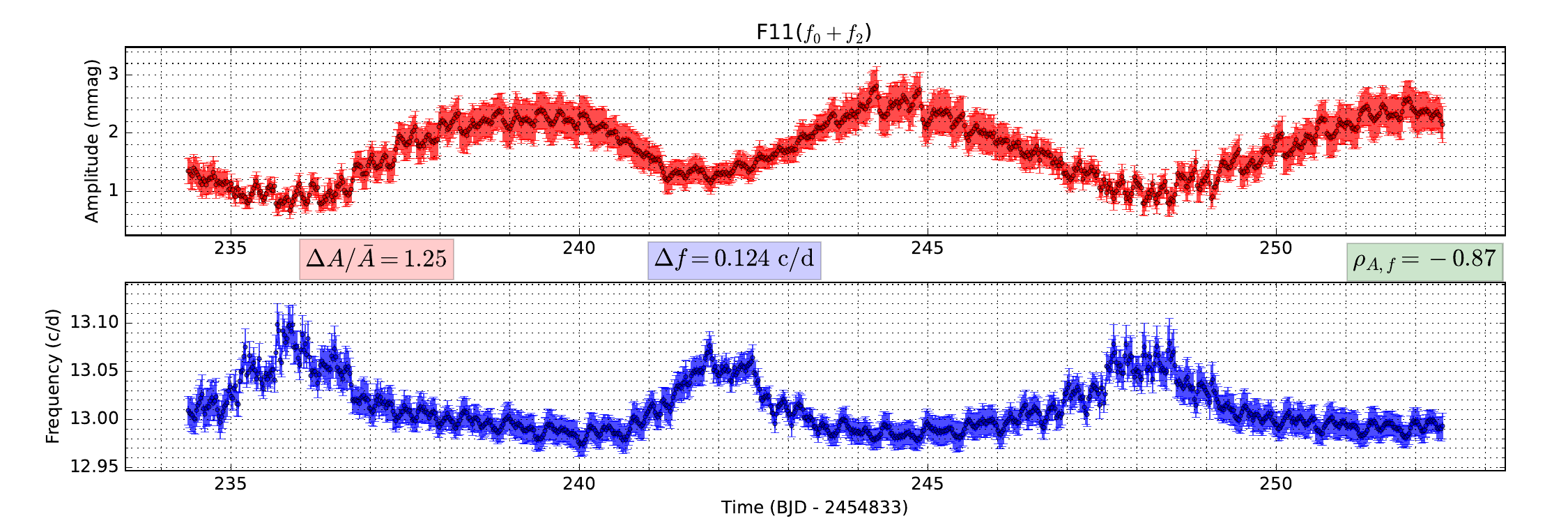}
  \includegraphics[width=0.8\textwidth]{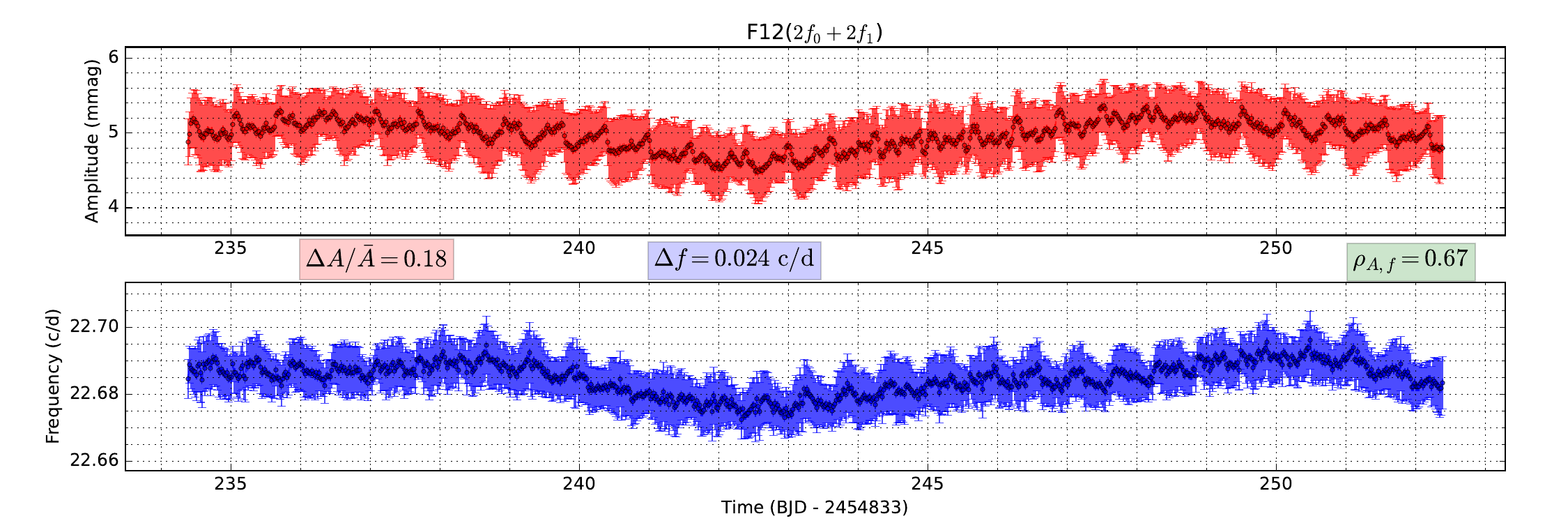}
  \includegraphics[width=0.8\textwidth]{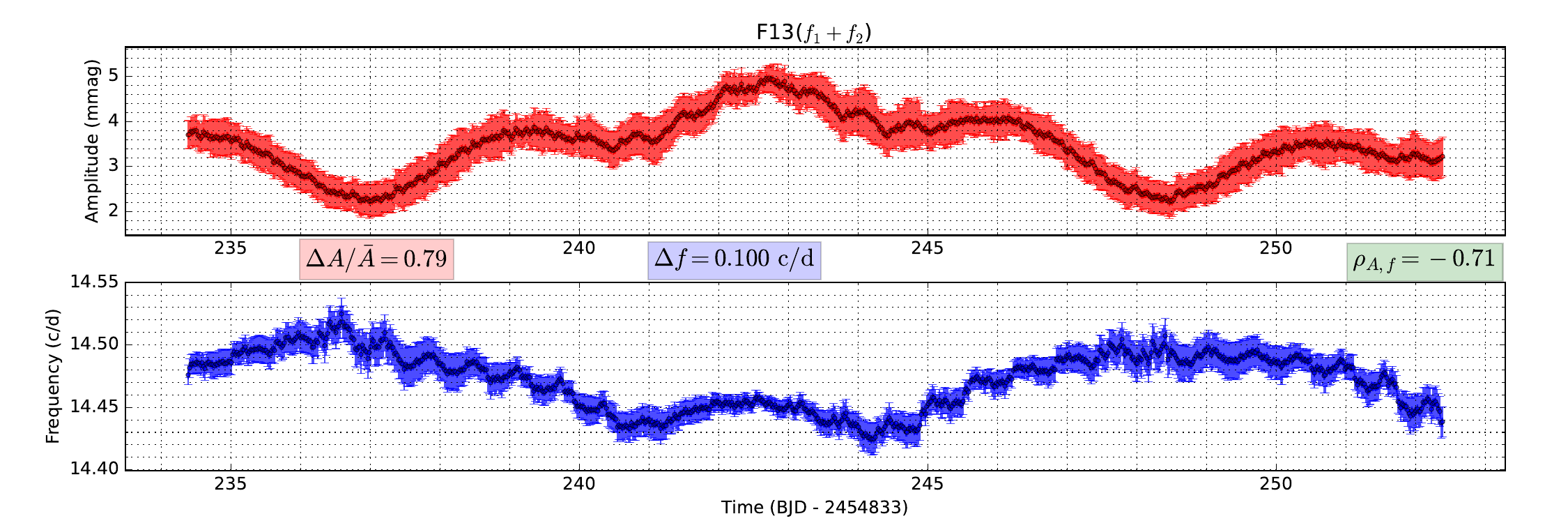}
  \includegraphics[width=0.8\textwidth]{fig/amp_freq_absvar_17.pdf}
  \caption{Amplitude and frequency variations of the 19 pulsation modes, Part III.}
  \label{fig:var_amp_freq03}
\end{figure*}

\begin{figure*}[htp]
  \centering
  \includegraphics[width=0.8\textwidth]{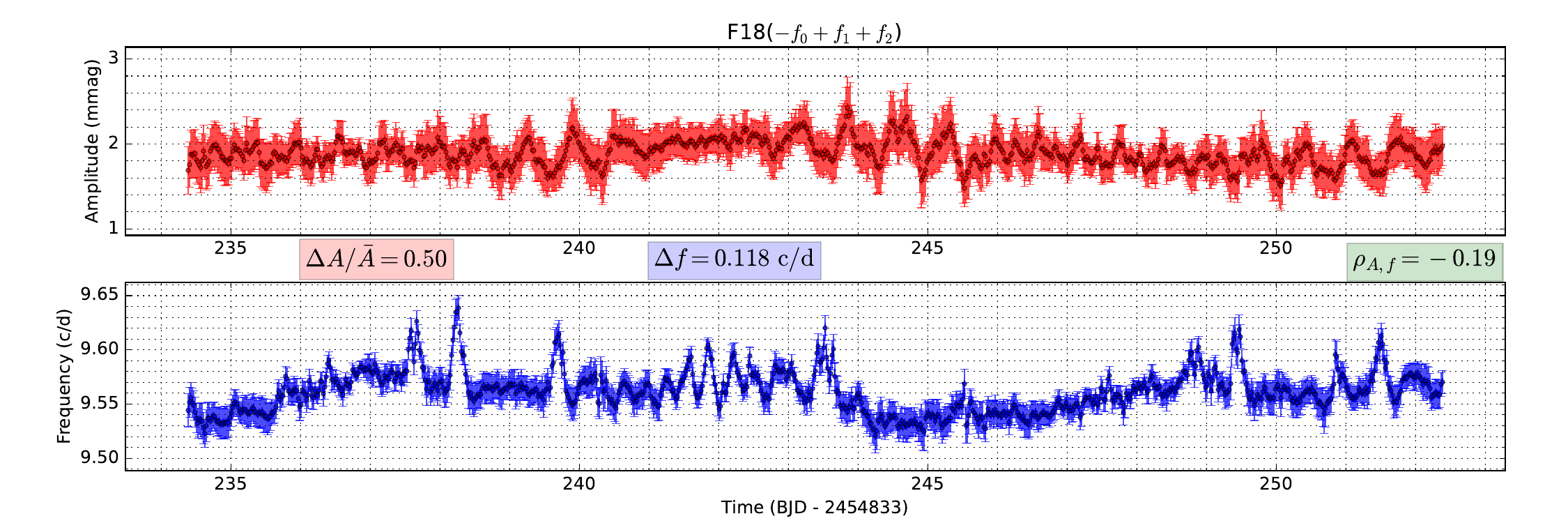}
  \includegraphics[width=0.8\textwidth]{fig/amp_freq_absvar_19.pdf}
  \includegraphics[width=0.8\textwidth]{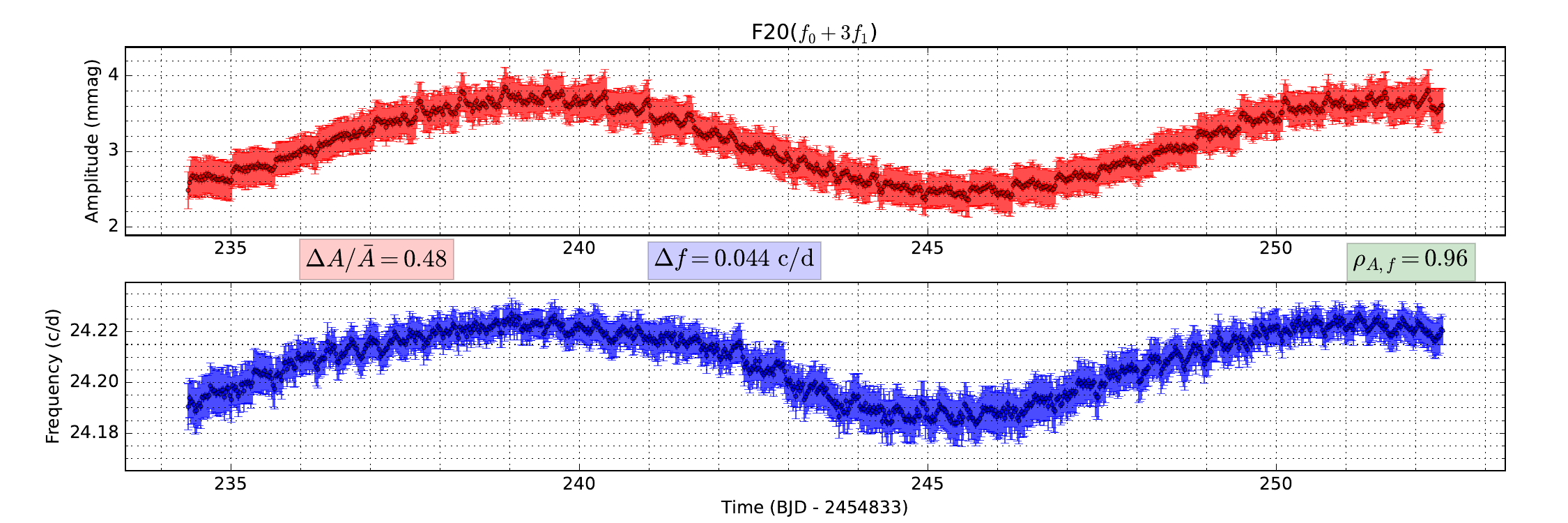}
  \includegraphics[width=0.8\textwidth]{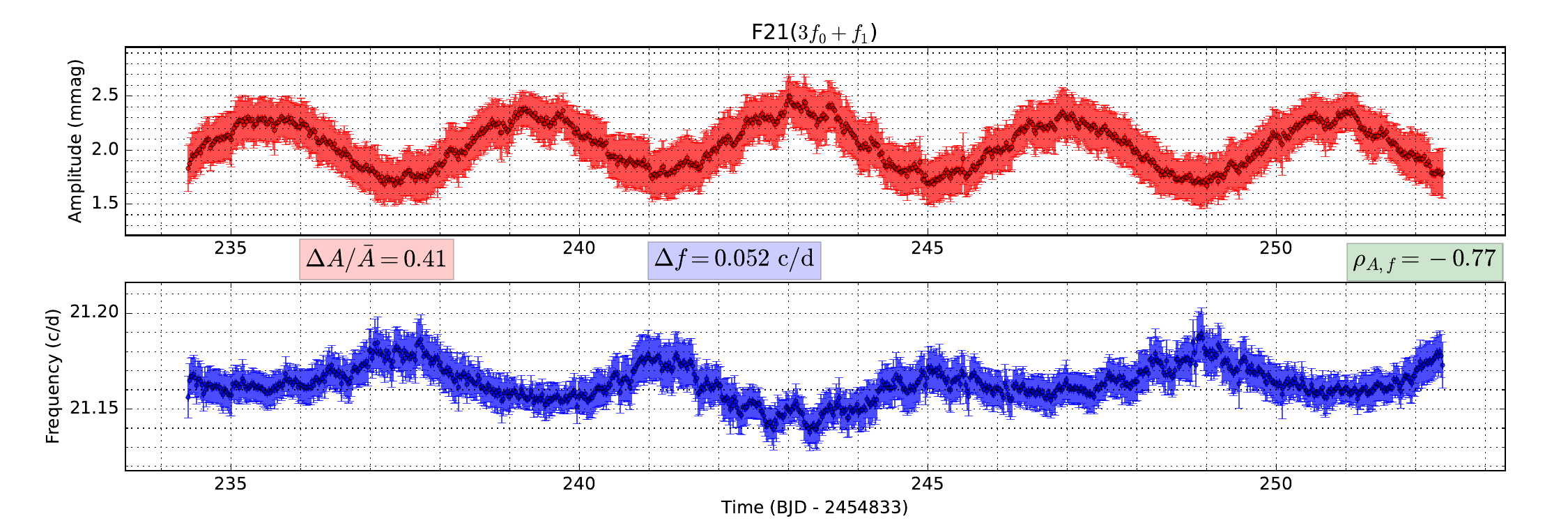}
  \caption{Amplitude and frequency variations of the 19 pulsation modes, Part IV.}
  \label{fig:var_amp_freq04}
\end{figure*}

Table \ref{tab:freq_solution} presents the absolute and relative variations of these 19 pulsation modes. The absolute variation is defined as the difference between the maximum and minimum values of the mode ($\Delta f \equiv f_\mathrm{max} - f_\mathrm{min}$ and $\Delta A \equiv A_\mathrm{max} - A_\mathrm{min}$), while the relative variation is the ratio of the difference to the mean values ($\Delta f/\bar{f} \equiv (f_\mathrm{max} - f_\mathrm{min})/\bar{f}$ and $\Delta A/\bar{A} \equiv (A_\mathrm{max} - A_\mathrm{min})/\bar{A}$).
The relative variations of the amplitudes and the absolute variations of the frequencies are also presented in Figures \ref{fig:var_amp_freq01}, \ref{fig:var_amp_freq02}, \ref{fig:var_amp_freq03}, and \ref{fig:var_amp_freq04}.

The $\rho_{A,f}$ in Table \ref{tab:freq_solution} and Figures \ref{fig:var_amp_freq01}, \ref{fig:var_amp_freq02}, \ref{fig:var_amp_freq03}, and \ref{fig:var_amp_freq04} represents the Pearson correlation coefficient between the amplitude and frequency for each of the 19 pulsation modes.

\subsection{Time-frequency diagram}

\begin{figure*}[htp]
  \centering
  \includegraphics[width=0.8\textwidth]{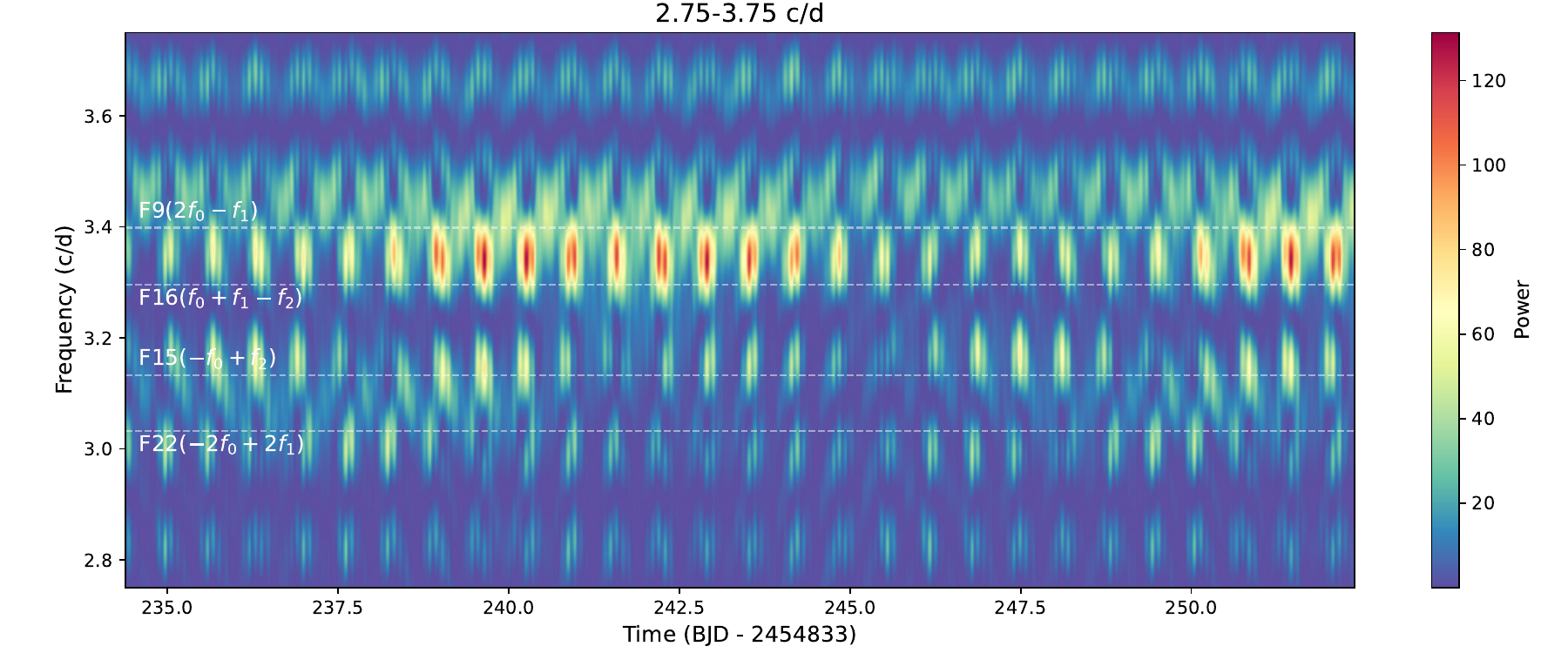}
  \caption{Time-frequency diagram which presents interaction details in $2-3\ \cd$.}
  \label{fig:var_2-3}
\end{figure*}

The time-frequency diagram (Subfigure (c) of Figure \ref{fig:var_f2} and Figure \ref{fig:var_2-3}) was plotted with a time window of 6 days, which was moved from the start to the end time of the SC data, with a step of 0.01 days. The power (square of amplitude) is indicated by the color bar.

Please note that the pulsation modes marked in the time-frequency diagram are for illustration purposes only and may not be accurate. This is because it is difficult to identify a pulsation mode with drastic variations in amplitude and frequency, let alone strong interaction with nearby modes.

\section{Figures and Tables}


\label{lastpage}

\end{CJK*}
\end{document}